\documentclass[sigconf,screen,nonacm]{acmart}

\AtBeginDocument{%
  }

\usepackage{listings}
\usepackage{xcolor}

\usepackage{microtype}
\usepackage{multirow}
\usepackage{enumitem}
\usepackage{cleveref}
\usepackage{hyperref}
\usepackage{xspace}

\lstdefinelanguage{dafny}{
  morekeywords={method,function,returns,requires,ensures,invariant,decreases,var,while,if,else,forall,assert,new,int,real,bool,array,seq,nat,true,false,modifies},
  sensitive=true,
  morecomment=[l]{//},
  morecomment=[s]{/*}{*/},
  morestring=[b]",
  basicstyle=\ttfamily\small,
  keywordstyle=\bfseries,
  commentstyle=\itshape\color{gray},
  numbers=left,
  numberstyle=\tiny\color{gray},
  numbersep=5pt,
  frame=single,
  breaklines=true,
  captionpos=b
}

\begin{document}

\balance

\title{ProofPulse: Interactive Proof Coverage Analysis for Dafny}

\author{Álvaro F. Silva}
\correspondingauthor
\orcid{0009-0005-2941-9942}
\affiliation{%
  \institution{INESC TEC, Faculdade de Engenharia, Universidade do Porto}
  \city{Porto}
  \country{Portugal}
}
\email{up201603524@fe.up.pt}

\author{Ruben Martins}
\orcid{0000-0003-1525-1382}
\affiliation{%
  \institution{Carnegie Mellon University}
  \department{Computer Science Department}
  \city{Pittsburgh}
  \country{USA}
}
\email{rubenm@andrew.cmu.edu}

\author{Alexandra Mendes}
\authornote{Author order follows a senior-author-last convention.}
\orcid{0000-0001-8060-5920}
\affiliation{%
  \institution{INESC TEC, Faculdade de Engenharia, Universidade do Porto}
  \city{Porto}
  \country{Portugal}
}
\email{alexandra@archimendes.com}

\renewcommand{\shortauthors}{Silva et al.}
\newcommand{\proofpulse}{\textbf{ProofPulse}\xspace}
\newcommand{\para}[1]{\par\smallskip\noindent\textbf{#1.}}
\newcommand{\parait}[1]{\par\smallskip\noindent\textit{#1.}}
\newcommand{\covtest}{{\ttfamily CovTest}\xspace}
\newcommand{\covcomplete}{{\ttfamily CovComplete}\xspace}
\newcommand{\uncovered}{{\ttfamily Uncovered}\xspace}

\begin{abstract}
Deductive verification ensures that an implementation satisfies its specification, but successful verification does not guarantee the quality of the specification. As such, weak specifications and redundant invariants may create overconfidence in ``verified'' code.

We present \proofpulse, a tool for Dafny that diagnoses specification quality using a three-valued proof coverage model. By analyzing proof dependencies, ProofPulse distinguishes between elements that contribute to specification intent, those used only for auxiliary checks, and those irrelevant to any proof obligation.

Evaluated against an oracle of 252 programs from the dafny-synthesis benchmark, ProofPulse provides a high-precision signal for specification weaknesses, particularly unnecessary preconditions and vacuous proofs. With unsat-core minimization, ProofPulse achieves perfect precision for precondition classification and reduces false positives across all evaluated categories. These results show that proof coverage is a practical complement to verification. Although it cannot fully capture semantic intent, it can reveal weak proof coupling in programs that otherwise appear fully verified.

Just as a pulse check distinguishes vitality from the mere absence of symptoms,  ProofPulse exposes weaknesses in proofs that technically verify successfully but still suffer from inadequate or redundant code and specifications. 

\noindent
\textbf{Demo:} \href{https://www.youtube.com/watch?v=8pO3NAodjoQ}{\ttfamily \small https://www.youtube.com/watch?v=8pO3NAodjoQ}\\
\textbf{Code:} \href{https://github.com/VeriFixer/ProofPulse}{\ttfamily \small https://github.com/VeriFixer/ProofPulse}\\
\textbf{Prebuilt Docker image:} \href{https://doi.org/10.5281/zenodo.21174686}{\ttfamily \small https://doi.org/10.5281/zenodo.21174686}
\end{abstract}



\maketitle

\section{Introduction}

Deductive software verification tools, such as Dafny~\cite{leino2010dafny}, ensure that implementations satisfy their specifications. However, a successful verification only proves that the implementation satisfies the \emph{given} specification, it does not ensure that the contract is adequately coupled to the code it governs. Weak postconditions, redundant invariants, or unnecessary preconditions may allow verification to succeed while large parts of the program remain unconstrained. In the extreme case, the Dafny verifier may report a ``correct'' status for a program that lacks verification conditions. In such instances, the IDE provides a passing ``tick'' despite the absence of meaningful relationship between the specification and the code, creating confidence in an implementation that remains effectively unverified.

Tomb and Joshi~\cite{tomb2025static} introduced \emph{verification coverage}, using unsatisfiable cores to identify which program elements are relevant to a proof. While effective, their approach is binary (covered/uncovered) and primarily exposed via command-line output.
Such output can be difficult to inspect, since developers must manually connect coverage information back to source locations and proof dependencies. This makes it hard to distinguish elements that establish the main specification from elements used only for auxiliary checks, such as bounds checks or intermediate assertions.

We present \textbf{ProofPulse}, a tool that makes proof coverage more interactive and informative, while integrating it into the IDE. Our key insight is that not all ``covered'' elements play the same role: some contribute directly to the specification, while others only support auxiliary checks. Our main \textbf{contributions} are:

\begin{enumerate}[leftmargin=*]
\item[\small{$\blacktriangleright$}]  \textbf{ProofPulse VSCode Extension}: An interactive tool (VSCode + web viewer) for exploring proof dependencies, designed for Dafny users across all levels of expertise.

\item[\small{$\blacktriangleright$}] \textbf{Three-valued coverage mapping}: A coverage model distinguishing essential, auxiliary, and irrelevant proof elements.

\item[\small{$\blacktriangleright$}] \textbf{Z3 unsat-core minimization}: An optional unsatisfiable core minimization layer over Z3 that improves attribution quality.

\item[\small{$\blacktriangleright$}] \textbf{Empirical Evaluation}: Evaluation on 252 programs showing
that ProofPulse provides a high-precision signal for unnecessary preconditions, vacuous proofs, and weak proof coupling.
\end{enumerate}

\section{Proof Coverage}

ProofPulse refines binary coverage into a three-valued proof-relevance model computed from Dafny's dependency information.

\para{From Binary to Three-Valued Coverage}
Tomb and Joshi~\cite{tomb2025static} define \emph{verification coverage} as a semantic property of proof relevance: a program element is \emph{covered} if modifying it may cause verification to fail, and \emph{uncovered} if it can be changed arbitrarily without affecting the proof. Their approach instruments Boogie's verification condition (VC) generation with labels on assertions and assumptions. After discharging the VC, the unsatisfiable core returned by the SMT solver identifies the covered elements.

In addition, they provide a semantic interpretation of \emph{uncovered} elements depending on their role in the verification process:

\begin{description}[leftmargin=!]
\item[Assignments or calls:] unconstrained or unreachable code.
\item[Assumptions:] unnecessary assumptions, including \texttt{assume}
statements or preconditions of the verified procedure.
\item[Callee preconditions:] potentially vacuous calls.
\item[Callee postconditions:] assumptions not required by the caller.
\item[Procedure postconditions:] vacuously proved guarantees.
\item[Assertions:] vacuous goals or unused intermediate proof steps.
\item[Loop invariants:] vacuous proofs when proved, or unnecessary assumptions
when used in the loop body.
\end{description}

ProofPulse adopts the same semantic view of proof relevance. Uncovered elements are those that do not contribute to any proof obligation, matching Tomb and Joshi's notion of irrelevance. However, ProofPulse refines their binary covered/uncovered classification by separating fully irrelevant elements from elements that are relevant only to auxiliary obligations.

Some elements are verified and used by the proof, but not to establish the main specification. For example, a callee specification may be checked or available at a call site without being needed to prove the caller's postcondition. Rather than reporting such elements as fully uncovered, ProofPulse marks them as \textsc{CovTest}, indicating that they support verification but are not essential to the primary proof objective.
As a result, ProofPulse preserves the core semantic insight of prior work while providing a more fine-grained interpretation of proof relevance.

\para{Coverage Computation}
ProofPulse computes coverage in two phases. First, it builds a proof dependency graph and assigns a uniform \emph{internal} coverage status to every node via breadth-first search propagation. Second, a type-aware \emph{refinement} step maps internal statuses to a final three-valued classification depending on each node's semantic role.

\parait{Proof Dependency Graph}  From a Dafny source file and verification log,
ProofPulse constructs a directed graph $G = (V, E, T)$ where:
\begin{itemize}[leftmargin=*]
\item Each node $v \in V$ corresponds to a source-level element identified by its file location span.
\item Each directed edge $(u, v) \in E$ indicates that $v$ was used to discharge the proof obligation of $u$.
\item $T \subseteq V$ is the set of \emph{top nodes}: the proof obligations that Dafny reports (postconditions, manual assertions, automatic assertions such as index-in-range checks).
\end{itemize}

Each node is classified by a \emph{token type} $\tau(v) \in \{\texttt{Postcondition}, \allowbreak \texttt{Precondition}, \allowbreak \texttt{AssertManual}, \allowbreak \texttt{AssertAuto}, \allowbreak \texttt{CodeLine}\}$, inferred \allowbreak from the proof message text. Every node starts as \uncovered. The algorithm then propagates coverage from top nodes downward through the dependency edges. After these passes, every node $v$ has an internal status $\sigma_{\mathit{int}}(v) \in \{\textrm{\covcomplete}, \textrm{\covtest}, \textrm{\uncovered}\}$. A node is marked \covcomplete when it participates in proving a postcondition, \covtest when it participates only in proving non-postcondition obligations, such as index safety or manual assertions, and \uncovered when no proof obligation depends on it. The graph can be navigated interactively in the ProofPulse web viewer.

\begin{figure}[t]
  \centering
  \includegraphics[width=0.85\linewidth]{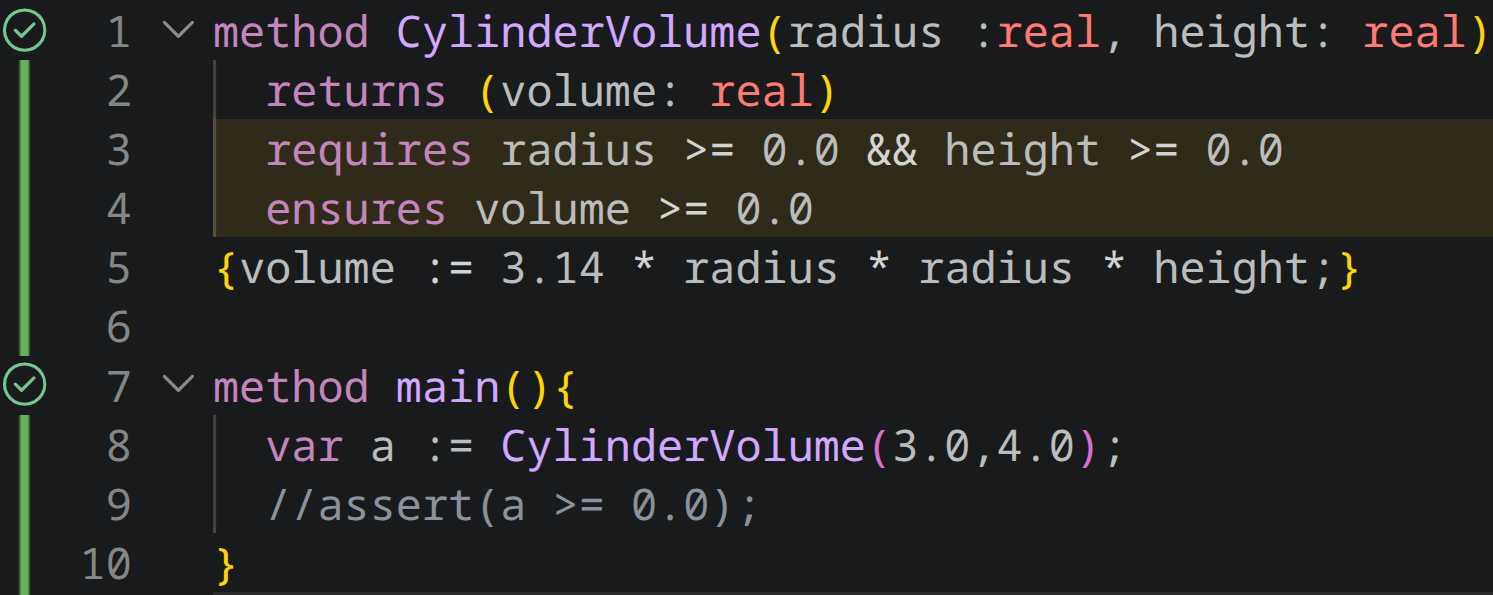}
  \caption{ProofPulse VSCode extension. The precondition \texttt{radius >= 0.0} is classified as \covtest (yellow highlight).%
  }
  \label{fig:vscode}
\end{figure}

\parait{Type-Aware Refinement}
Statuses are further refined based on the element type. The mapping resembles the one proposed by Tomb and Joshi, differing relative to the \covtest cases:

\begin{enumerate}[leftmargin=*]
\item When a \emph{postcondition} is never needed by any callee, either because the method is never called or because the postcondition is not used to prove any property in the callee, we mark it as \covtest instead of \uncovered. This distinguishes it from the stronger case in which the postcondition is proved vacuously without relying on any code line or lemma in the method body. In the example of ~\Cref{fig:vscode}, uncommenting line~9 would cause the postcondition to be classified as \covcomplete.

\item A \emph{precondition} is marked as \covtest if it is used by a callee, but is not strictly necessary to prove the method's postcondition. This distinction is useful because some preconditions exist primarily to constrain caller behavior rather than to serve as requirements for establishing the postcondition, as illustrated in \Cref{fig:vscode}. If line~8 were commented out, the precondition would instead be classified as \uncovered, since it would neither be exercised by callers nor required to prove the postcondition.
\end{enumerate}

The introduction of \covtest addresses cases that would otherwise be misleadingly classified as either \uncovered or \covcomplete. In \Cref{fig:vscode}, the precondition \texttt{radius >= 0.0} is classified as \covtest because it is exercised by the caller at line~8, yet it is unnecessary for proving the postcondition, since \texttt{radius * radius} is always non-negative. Marking it as \uncovered would incorrectly suggest that the precondition is useless, despite its important semantic role in constraining caller behavior (i.e., in preventing negative radius values), while marking it as \covcomplete would hide the fact that it could potentially be weakened or removed without affecting the proof. Thus, \covtest acts as an intermediate warning category, highlighting specifications that participate in verification but may deserve further review or simplification.

\begin{figure*}[t]
  \centering
  \includegraphics[width=0.85\textwidth]{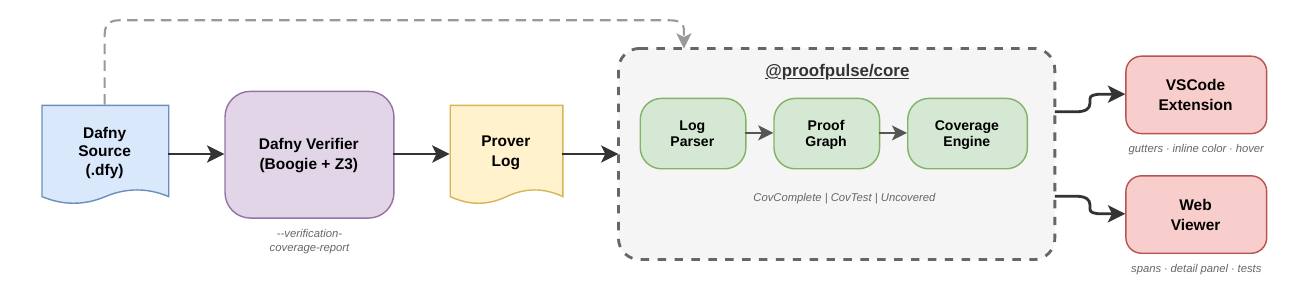}
  \caption{ProofPulse pipeline. The Dafny verifier produces a prover log with proof dependencies. The core library parses this into a proof dependency graph, computes three-valued coverage, and feeds two visualization front-ends.}
  \Description{Architecture diagram showing the four-stage ProofPulse pipeline from Dafny source to visualization.}
  \label{fig:pipeline}
\end{figure*}

\section{\proofpulse Architecture}
\label{sec:architecture}

ProofPulse is implemented in TypeScript with two packages: a shared core library (\texttt{@proofpulse/core}) and a VSCode extension. The pipeline, presented in \Cref{fig:pipeline}, operates in four stages:

\para{1. Verification with coverage logging}
ProofPulse invokes Dafny with the \texttt{-{}-verification-coverage-report} flag, which enables the coverage instrumentation described by Tomb and Joshi~\cite{tomb2025static}. Dafny then produces a prover log with proof dependencies.

\para{2. Log parsing and graph construction}
The core library parses the prover log and constructs the proof dependency graph. Each source span becomes a node and dependency relationships become directed edges.

\para{3. Coverage computation}
The graph is traversed to assign coverage statuses. The algorithm processes top-level obligations first, then propagates status through the dependency edges.

\para{4. Visualization}
Two front-ends consume the proof graph:
\begin{itemize}[leftmargin=*,topsep=0pt]
\item \emph{VSCode extension}: Provides gutter decorations (red for \uncovered, yellow for \covtest), and hover diagnostics showing the proof message and coverage status (\Cref{fig:vscode}).
\item \emph{Web viewer}: A two-column layout with a tokenized source editor with clickable spans, and a detail panel for exploring dependencies. It can be activated through the VSCode extension (\Cref{fig:webviewer}).
\end{itemize}

\section{Evaluation}
\label{sec:evaluation}

We evaluate whether proof coverage is a useful practical proxy for specification quality by comparing ProofPulse against the manually curated oracle from the \texttt{dafny-synthesis} benchmark~\cite{misu2024towards}, which contains 252 verified Dafny programs.
Proof coverage and semantic specification strength are related but distinct notions. A specification is semantically weak when it under-specifies intended behavior. This weakness often manifests itself as weak proof coupling, where specifications, invariants, or code do not contribute meaningfully to verification. Although coverage cannot determine whether a specification fully captures program intent, it can reveal redundancy, vacuity, and under-constrained proofs. We therefore evaluate how well coverage signals align with human judgments of specification quality.

\subsection{Benchmark and Oracle}

The \texttt{dafny-synthesis} benchmark~\cite{misu2024towards} contains Dafny programs generated by GPT-4 and PaLM-2 under multiple prompting strategies. In Misu et al. ~\cite{misu2024towards}, each verified program is manually annotated with labels for postcondition strength (\textsc{Strong}/\textsc{Weak}/\textsc{Wrong}), precondition necessity (\textsc{Required}/\textsc{Optional}), and loop invariant strength (\textsc{Strong} / \textsc{Weak}). 
These oracle labels are semantic judgments, whereas ProofPulse measures proof coupling. For example, an oracle-\textsc{Strong} postcondition captures intended behavior, while a covered postcondition only means that the proof depends on it.
We map ProofPulse's statuses to Oracle's categories as follows:

\begin{itemize}[leftmargin=*,topsep=0pt]
\item \emph{Postconditions}: \textsc{Strong} if all postconditions are covered
      (\covtest or \covcomplete)
      and all body code lines are \covcomplete;
      \textsc{Weak} otherwise. If parts of the code are not required for proving postconditions, this indicates either under-constrained specifications or irrelevant code.

\item \emph{Preconditions}: \textsc{Required} if any precondition is covered;
      \textsc{Optional} if all preconditions are \uncovered. Uncovered preconditions can be safely removed. Hence, they are \textsc{Optional}.
      
\item \emph{Invariants}: \textsc{Strong} if all loop invariant nodes are covered;
      \textsc{Weak} if any is \uncovered. Same reasoning as postconditions.

\end{itemize}

\subsection{Results}

\para{Effect of Unsat-Core Minimization} During the development of ProofPulse, we observed that non-minimal Z3 unsatisfiable cores can affect proof-coverage attribution by introducing unnecessary dependencies. Tomb and Joshi~\cite{tomb2025static} similarly note that SMT solvers do not guarantee minimal unsat cores. To mitigate this issue, ProofPulse optionally applies deletion-based unsat-core minimization~\cite{marques2011improving} before coverage classification. As shown in \Cref{tab:overall-results}, minimization reduces false positives across all categories (e.g., from 24 to  21 for postconditions) and consistently improves precision, while preserving recall. 
In our benchmark, the minimized cores were small enough that minimization added little overhead while remaining practical for interactive use.

\para{Result Analysis} As shown in \Cref{tab:overall-results}, ProofPulse performs particularly well for preconditions, achieving perfect precision and  0.96 accuracy with core minimization. This aligns with the underlying semantics: an uncovered precondition is genuinely unnecessary for verification, closely matching the oracle’s \textsc{Optional} label. 

\begin{table}[t]
  \centering
  \caption{Comparison of ProofPulse classification performance of the base configuration (Base) and core minimization (Min). The table reports confusion-matrix counts (True Positive - TP, False Positive - FP, False Negative - FN, True Negative - TN) together with precision (Prec), recall (Rec), and accuracy (Acc)}
  \label{tab:overall-results}

  \begin{tabular}{lcrrrrrrr}
    \toprule
    \textbf{Cat.} & \textbf{Cfg} & \textbf{TP} & \textbf{FP} & \textbf{FN} & \textbf{TN} & \textbf{Prec} & \textbf{Rec} & \textbf{Acc} \\
    \midrule

    \multirow{2}{*}{Post.}
      & Base & 184 & 24 & 0 & 11 & 0.88 & 1.00 & 0.89 \\
      & Min  & 184 & 21 & 0 & 13 & 0.90 & 1.00 & 0.90 \\

    \midrule

    \multirow{2}{*}{Pre.}
      & Base & 63 & 3 & 7 & 96 & 0.95 & 0.90 & 0.94 \\
      & Min  & 63 & 0 & 6 & 98 & 1.00 & 0.91 & 0.96 \\

    \midrule

    \multirow{2}{*}{Inv.}
      & Base & 77 & 13 & 0 & 11 & 0.86 & 1.00 & 0.87 \\
      & Min  & 77 & 10 & 0 & 13 & 0.89 & 1.00 & 0.90 \\

    \bottomrule
  \end{tabular}
\end{table}

\begin{figure*}[t]
  \centering
  \includegraphics[width=0.65\textwidth]{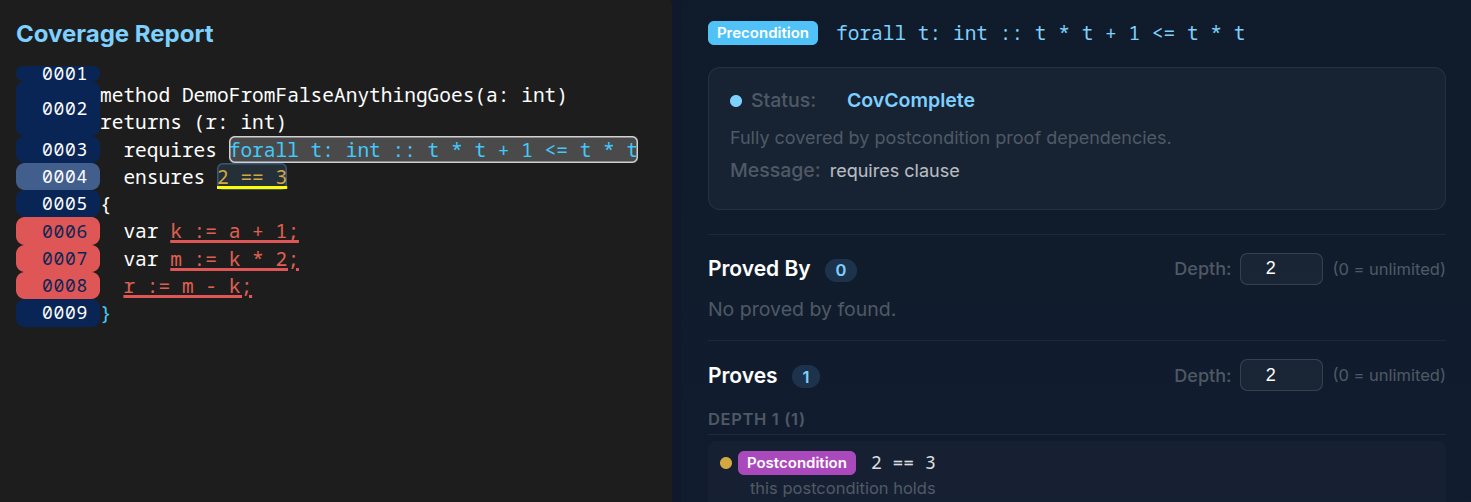}
  \caption{ProofPulse web viewer showing a Dafny method with an unsatisfiable precondition. On the left, the code is displayed with red underlining indicating \uncovered elements, yellow highlighting indicating \covtest, and blue highlighting indicating the currently selected item. On the right, more detailed information about the selected item is shown, including its status, proof dependencies, and the elements that prove it.
  }
  \label{fig:webviewer}
\end{figure*}

Postcondition precision is also high ( 0.90), indicating that ProofPulse's coverage-based \textsc{Strong} classifications usually agree with the oracle. Precision stays below 1.00 because a postcondition can be fully covered, with every line exercised, and still be weak, in which case ProofPulse classifies it as Strong while the oracle labels it weak. These fully-covered-but-weak postconditions are the false positives that lower precision. \Cref{fig:vscode} illustrates one such case, where every line of \texttt{CylinderVolume} is covered and exercised, yet the postcondition remains weak. Postcondition recall is perfect (1.00). When ProofPulse classifies a postcondition as weak, it is almost certainly weak. A weak flag means that some behavior of the implementation is left unconstrained by the postcondition. The exception is dead code: if unreachable code contains behavior the postcondition does not pin down, ProofPulse still flags the postcondition as weak, even though the specification could be strong, since that code can never execute. Such a case would be a false negative and would lower recall. In our evaluation no such cases occurred, so no truly strong postcondition was ever flagged as weak, leading to a perfect recall.

Invariant accuracy %
closely follows the behavior observed for postcondition classification. This is expected, since both invariants and postconditions constrain program behavior in a similar way. Loop invariants additionally have a dual role: they act as both preconditions and postconditions at different program points.

ProofPulse additionally detects vacuous proofs arising from contradictory assumptions. Figure~\ref{fig:webviewer} shows a method with an unsatisfiable \texttt{forall} precondition, where the method body is marked as \uncovered (marked in red). The web viewer’s proof graph further reveals that the contradictory \texttt{requires} clause alone suffices to discharge the postcondition (\texttt{2 == 3}), exposing the vacuous reasoning.

\para{Limitations}
ProofPulse's main limitation is imprecision in the Dafny-to-Boogie-to-Z3 pipeline, especially the lossy mapping from Boogie IVL back to Dafny source spans. The major problem is:
\begin{itemize}[leftmargin=*]

\item \emph{Quantified expressions (\texttt{forall}).} Dafny encodes quantified specifications as Boogie axioms with SMT triggers. These triggers are not part of the core logical proof and do not appear in Z3’s unsat core, so the corresponding source lines are always reported as uncovered, even when they are necessary for verification\,---\,reflecting a limitation in quantifier coverage tracking.

\end{itemize}

These are engineering limitations of the Dafny/Boogie pipeline rather than the coverage approach itself. Improved source tracking would directly improve ProofPulse.

\section{Related Work}

\para{Verification coverage}
Tomb and Joshi~\cite{tomb2025static} formalize static coverage for deductive verification using unsatisfiable cores, building on the coverage--vacuity duality from model checking~\cite{kupferman2008theory}. ProofPulse builds directly on their Boogie/Dafny implementation, extending it with a three-valued formalism, a proof dependency graph, and interactive tooling. Ghassabani et al.~\cite{ghassabani2016efficient, ghassabani2017proof} describe inductive validity cores for unbounded model checking and use them to measure implementation coverage. 

\para{Specification quality}
Le et al.~\cite{le2018verification} combine proof and test coverage using mutation-based metrics. Smoke testing~\cite{tomb2025static} inserts \texttt{assert false} at program points to detect vacuity, but produces many warnings and does not identify redundant specifications. ProofPulse provides finer-grained feedback with fewer false positives.

\para{Autoformalization evaluation}
Misu et al.~\cite{misu2024towards} evaluate LLM-generated Dafny specifications and provide the oracle used by us. Their manual classification of specification strength motivates automated approaches like ProofPulse for scalable quality assessment.

\para{Verification IDEs}
Dafny's built-in IDE support provides verification status (verified/error) but does not expose coverage information. The Dafny VSCode extension shows verification errors inline. ProofPulse complements this by showing \emph{what the proof used}, not just whether it succeeded.

\para{Usability} Oliveira et al.~\cite{oliveiraChallenges} studied the challenges practitioners face when using verification-aware languages. Their findings highlight the need for improved usability feedback and greater access to the internal reasoning of verification tools. The design of ProofPulse is motivated by these challenges, providing detailed verification feedback and exposing proof dependencies in an accessible way.

\section{Conclusion and Future Work}
We presented ProofPulse, a tool for interactive proof coverage analysis of Dafny programs. By constructing a proof dependency graph and applying a three-valued coverage formalism, ProofPulse provides actionable feedback about specification quality directly within the developer’s IDE. We also extend prior work with an optional deletion-based unsat-core minimization, which improves attribution precision by reducing spurious proof dependencies introduced by non-minimal SMT unsat cores.

Our evaluation on 252 LLM-generated Dafny programs demonstrates perfect precision for precondition classification and promising results for postconditions and invariants. Enabling unsat-core minimization further reduced false positives across all evaluated categories while preserving recall. The remaining limitations can largely be traced to Boogie-to-Dafny attribution gaps rather than fundamental issues with the approach itself.

Future work includes: (1) finer-grained sub-expression-level attribution, (2) user studies measuring the impact on proof debugging time and specification understanding, (3) extension to other verification-aware languages, and (4) improved quantifier handling.

\begin{acks}

Alexandra Mendes was partially funded by National Funds through the FCT - Fundação para a Ciência e a Tecnologia, I.P. (Portuguese Foundation for Science
and Technology) within the project VeriFixer, with reference 2023.15557.PEX (DOI: 10.54499/2023.15557.PEX) and by an Amazon Research Award, Fall 2024. 

Ruben Martins was partially supported by the National Science
Foundation (NSF) under Award CCF2427581 and DARPA Agreement FA8750-24-9-1000.

Álvaro Silva was co-financed by national funds through FCT – Fundação para a Ciência e a Tecnologia, I.P., under the support UID/50014/2025 (https://doi.org/10.54499/UID/50014/2025), by National Funds through the FCT - Fundação para a Ciência e a Tecnologia, I.P. (Portuguese Foundation for Science
and Technology) within the project VeriFixer, with reference 2023.15557.PEX (DOI: 10.54499/2023.15557.PEX), and Fundação para a Ciência e a Tecnologia (Portuguese Foundation for Science and Technology) through the Carnegie Mellon Portugal Program under the fellowship reference PRT/BD/155045/2024.

\end{acks}

\section{Data Availability Statement}

The data associated with this work are publicly available through a Zenodo
repository and can be accessed via the following DOI: \url{https://doi.org/10.5281/zenodo.21174686} \cite{alvaro_2026_21174686}

\bibliographystyle{ACM-Reference-Format}
\bibliography{software}

\end{document}